\documentclass[twoside]{sfm}

\input{sf.def}

\begin{document}

\thispagestyle{titlehead}

\setcounter{section}{0}
\setcounter{figure}{0}
\setcounter{table}{0}

\markboth{Korhonen}{Element spots in HgMn stars}

\titl{Element spots in HgMn stars}{Korhonen H.$^{1,2,3}$}
 {$^1$Centre for Star and Planet Formation, Natural History Museum of Denmark, 
  University of Copenhagen, {\O}ster Voldgade 5-7, DK-1350 Copenhagen, Denmark,
 email: {\tt heidi.korhonen@nbi.ku.dk} \\
 $^2$Finnish Centre for Astronomy with ESO (FINCA), University of 
  Turku, V{\"a}is{\"a}l{\"a}ntie 20, FI-21500 Piikki{\"o}, Finland\\
 $^3$Niels Bohr Institute, University of Copenhagen, Juliane Maries Vej 30, 
  DK-2100 Copenhagen, Denmark}

\abstre{A fraction of late B-type stars, the so-called HgMn stars, exhibit enhanced absorption lines of certain chemical elements, notably Hg and Mn, combined with an underabundance of He. For about a decade now the elements with anomalously high abundances in HgMn stars are known to be distributed inhomogeneously over the stellar surface. Temporal evolution of these elemental spots have been reported in a few HgMn stars, first secular evolution of the mercury spots in $\alpha$~And, and recently also a fast evolution of yttrium and strontium spots in HD~11753. The fast evolution of spots in HD 11753 is combined with a slower change in the overall abundance of the affected elements. In this paper I review what is known of elemental spots in HgMn stars and their secular and fast temporal evolution. 
}

\baselineskip 12pt

\section{Introduction}

A small group of late B-type stars show extreme overabundance of mercury and often also of manganese. Due to this chemical peculiarity these stars have been named HgMn stars. Unlike many other chemically peculiar stars, HgMn stars do not show enhancement of Rare Earth Elements. Although, heavy elements, like W, Re, Os, Ir, Pt, Au, Tl, Pb, and Bi, are often enhanced (see, e.g., \cite{Wahlgren95}). They can also exhibit anomalous isotopic abundances of, e.g., He, Hg, Pt, Tl, Pb, and Ca. (e.g., \cite{WoolfLambert99}, \cite{Dolk03}, \cite{CastelliHubrig04}, \cite{Cowley10}).

About 150 HgMn stars are currently known \cite{RensonManfroid09}. They are characterised by slower than normal rotation for their spectral type, and number of HgMn stars decreases with increasing rotational velocity \cite{WolfWolf74}. Additionally, they are found to preferably occur in binary and multiple systems. More than two thirds of the HgMn stars are known to belong to spectroscopic binaries \cite{HubrigMathys95}, with a preference of orbital periods ranging from 3 to 20 days. In a recent study Sch{\"o}ller et al. \cite{Schoeller10} observed 57 HgMn stars with the adaptive optics assisted imaging using ESO's Very Large Telescope and discovered that the sample comprised of 25 binaries, 3 triples and 1 quadruple. Nine of these companion candidates were found for the first time. Only five stars in the sample show no indication of multiplicity, taking into account that 44 systems are confirmed or suspected spectroscopic binaries.

Vertical stratification of some of the chemical elements is commonly seen in magnetic Ap stars and rapidly oscillating chemically peculiar (roAp) stars (e.g., \cite{Ryabchikova04}, \cite{Kochukhov06}, \cite{Shulyak09}). Some evidence of vertical stratification has also been seen in HgMn stars, e.g., of Cr \cite{SavanovHubrig03}, of Mn (\cite{Alecian82}, \cite{Sigut01}), and of Ga \cite{Lanz93}. On the other hand, the sample of Thiam et al. \cite{Thiam10} consisting of four HgMn stars showed stratification only in the case of HD~178065, and even then only in manganese. In addition Makaganiuk et al. \cite{Makaganiuk12} found no convincing evidence of stratification of Ti and Y in HD~11753.

In the following the chemical spots and their temporal evolution are discussed in detail. In the end the question whether or not HgMn stars also exhibit weak magnetic fields is briefly touched.

\section{First detections of chemical spots on HgMn stars}

The existence and non-existence of chemical spots on the surface of HgMn stars were discussed back and worth in the literature for three decades. These investigations used both photometry (e.g., \cite{Stift73}, \cite{Adelman94}) and spectroscopy (e.g., \cite{Takeda79}, \cite{Rakos81}, \cite{Zverko97}). Still, most of the photometric data were not conclusive and the obtained spectra were not of sufficiently high resolution and signal-to-noise ratio.

\begin{figure}[!t]
\begin{center}
\hbox{
 \includegraphics[width=8cm]{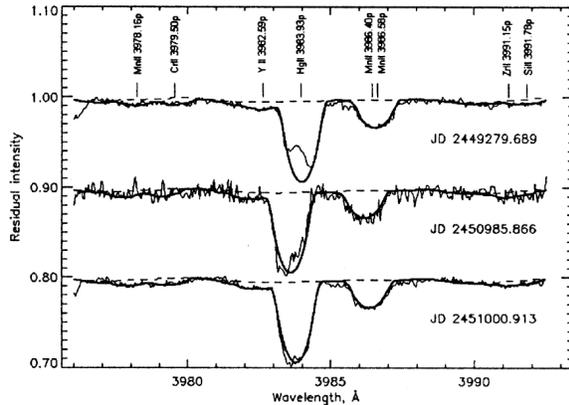}
}
\vspace{-5mm}
\caption[]{Observed spectra of $\alpha$~And obtained at three different time points and shown together with the computed binary spectra. Clear indications of variability in Hg profiles is seen. (from \cite{Ryabchikova99})}
\label{alphaAnd}
\end{center}
\end{figure}

The first more concrete evidence of chemical surface spots on HgMn stars gave from high resolution ($\lambda/\Delta \lambda$ approximately 40,000), high signal-to-noise ratio ($>$200) observations of $\alpha$~Andromedae analysed by Ryabchikova et al. \cite{Ryabchikova99}. They did a detailed chemical abundance analysis and at the same time discovered that the line profile of HgII 3983 {\AA} was changing between the observations obtained at different times (see Figure\,\ref{alphaAnd}). Unfortunately, they only had three epochs of observations and could not make firm conclusions about the changes. They write in their paper: "The most natural explanation of the observed profile variations is an inhomogeneous mercury distribution over the stellar surface. If it is supported by further observations then it will be the first indication for a spotty structure in HgMn stars" \cite{Ryabchikova99}.

Inspired by the possible detection of line profile variations in $\alpha$~And Adelman et al. \cite{Adelman02} carried out new high resolution observations of this intriguing target. With these new observations they achieved the first definitive detection of spectral variation in a HgMn star, that could not be attributed to the orbital motion of the companion. The variation they detect is produced by a non-uniform surface distribution of mercury that is concentrated in the equatorial region of the star. Similar variability was also reported by Wahlgren et al. \cite{Wahlgren01}

Since the first detections of chemical spots on HgMn stars several papers concentrating on them have been published (e.g., \cite{Kochukhov05}, \cite{Hubrig06a}, \cite{Briquet10}) and even maps of the surface element distributions have been published on handful of HgMn stars. The published surface maps are summarised in Table~\ref{HgMn_maps}, which gives the name of the target, elements for which maps have been obtained, from which year the observations are and where the map was published.

\begin{table}[!t]
\centering
\caption{HgMn stars with surface maps.} \label{HgMn_maps} \tabcolsep1.2mm
\begin{tabular}{llll}
\hline\noalign{\smallskip}
 Target      & Mapped elements & Year & Reference\\
\hline \noalign{\smallskip}
$\alpha$~And & Hg & 1998, 2002, 2004 & Kochukhov et al. \cite{Kochukhov07}\\
AR~Aur       & Hg, Y, Sr & 2005 & Savanov et al. \cite{Savanov09}\\
             & Y, Fe & 2005, 2008 & Hubrig et al. \cite{Hubrig10}\\
HD~11753     & Y, Sr, Ti & 2000 & Briquet et al. \cite{Briquet10}\\
             & Y, Sr, Ti, Cr & 2010 & Makaganiuk et al. \cite{Makaganiuk12}\\
             & Y, Sr, Ti, Cr & 2000, 2009 2010 & Korhonen et al. \cite{Korhonen13}\\
66~Eri       & Y, Sr, Ti, Ba & 2010 & Makaganiuk et al. \cite{Makaganiuk11a}\\
\noalign{\smallskip}\hline
\end{tabular}
\end{table}

\section{Spot locations in the binary reference frame}

The analysis of Sr and Y concentrations on the surface of AR~Aur by Hubrig et al. \cite{Hubrig06a} showed that these elements were most likely concentrated in the equatorial region of the star in a fractured ring-like structure. Hubrig et al. \cite{Hubrig06a} also interestingly noted that, in this tidally locked binary system, a large section of the elemental ring was missing on the area facing the secondary star, and a smaller section on almost the opposite side. These results were confirmed by new observations of AR~Aur published by Hubrig et al. \cite{Hubrig10}.

\begin{figure}[!t]
\begin{center}
\hbox{
 \includegraphics[width=11cm]{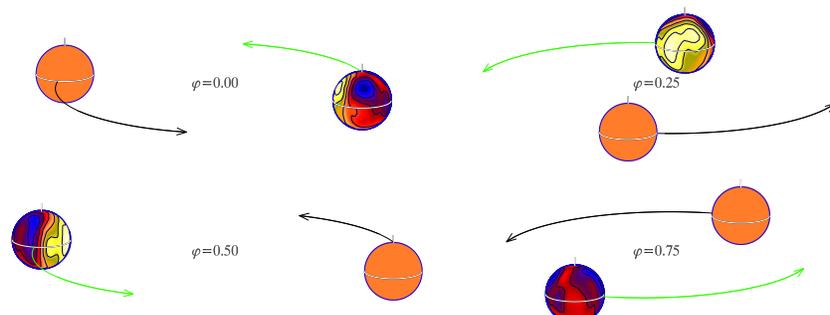}
}
\vspace{-5mm}
\caption[]{Orbital motion of 66~Eri binary and the Y distribution on the primary (darker colour indicates higher Y abundance). The secondary is marked with uniform colour. (from \cite{Makaganiuk11a})}
\label{66Eri}
\end{center}
\end{figure}

Similarly, Makaganiuk et al. \cite{Makaganiuk11a} report that in 66 Eri the elements Y, Sr, Ti, and Ba show a relative underabundance on the hemisphere facing the secondary star. Figure\,\ref{66Eri} shows the orbital motion of the primary and the Y abundance map of the primary. Makaganiuk et al. \cite{Makaganiuk11a} discuss that the reason for this underabundance could be a) that tidal effects cause the destruction of the upper layer of the atmosphere that contains the elemental excess, or b) the radiation from the secondary could impact the elemental distribution.

\section{Secular spot evolution}

The surface spots on magnetic Ap stars are very stable structures. Therefore it came as a surprise when Kochukhov et al \cite{Kochukhov07} reported slow variations in the surface configuration of mercury on the HgMn star $\alpha$~Andromedae. These secular variations were discovered using three datasets spanning in total seven years. Kochukhov et al. \cite{Kochukhov07} explained the behaviour with a non-equilibrium, dynamical evolution of the heavy-element clouds created by atomic diffusion, caused possibly by the same physical mechanism as the weather patterns on the planets.

Secular evolution of the elemental spots has also been reported on AR Aur \cite{Hubrig10}. The observations used in the study by Hubrig et al. \cite{Hubrig10} were obtained at a similar rotational phase and during three different years (2002, 2005, 2008). Still, they show clearly different line-profile shapes for Sr~II 4215.5~{\AA}, indicating secular evolution of the Sr spots.

Recent results on HD~11753 by Korhonen et al. \cite{Korhonen13} show that the over-all abundance of some of the elements has changed during the almost decade the observations span. Two out of the four elements which show rotational modulation, Y and Sr, also show a decrease in the over-all abundance between 2000 and 2009. Possibly still a further decrease occurred between August 2009 and January 2010.

\section{Fast spot evolution}

Discovering secular evolution of chemical spots in HgMn stars was surprising, but few years later changes in the chemical spots on even shorter time scales were discovered. Briquet et al. \cite{Briquet10} discovered that Y~II, Sr~II and Ti~II spots change configuration on HD~11753 between two datasets. These datasets were separated by approximately two months and clear evidence of dynamical spot evolution on such a short time scale was seen. 

Korhonen et al. \cite{Korhonen13} investigated further the reliability of the reported fast dynamical spot evolution on HD~11753. They confirmed, using both chemical spot maps and equivalent width measurements, that the exact spot configuration changed on time scale of months. The main spots remained at similar locations for over a decade, though. Currently, HD~11753 is the only HgMn star with reported fast evolution of the surface inhomogeneities.

\section{Magnetic fields in HgMn stars}

Many spotted Ap stars show strong organised magnetic fields (e.g., \cite{Silvester12}). On the other hand, the issue of whether HgMn stars exhibit magnetic fields is still hotly debated in the literature. Detections of magnetic fields in some HgMn stars have been reported over the years (e.g., \cite{MathysHubrig95}, \cite{Hubrig99}, \cite{HubrigCastelli01}, \cite{Hubrig12}). On the other hand, other studies have not found magnetic fields in HgMn stars (e.g., \cite{Shorlin02}, \cite{Wade06}, \cite{Makaganiuk11b}, \cite{Kochukhov13}).

At times the results from different groups agree, like in the case of the non-detection of magnetic field in HD 71066 (\cite{Hubrig99}, \cite{Hubrig06b}, \cite{Kochukhov13}). But there are other instances when the results do not agree, e.g., Hubrig et al. \cite{Hubrig10} detect weak field in AR~Aur whereas Folsom et al. \cite{Folsom10} do not. In the cases of discrepancy most of the non-detections are obtained using the LSD method, where information on many different lines is combined to improve signal-to-noise-ration, and the detections are typically made using the moment technique.

\begin{figure}[!t]
\begin{center}
\hbox{
 \includegraphics[width=11cm]{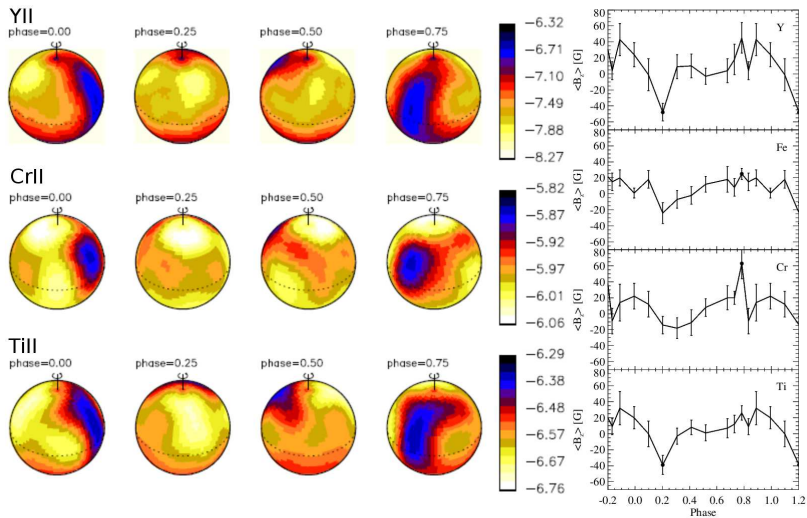}
}
\vspace{-5mm}
\caption[]{{\bf Left} Results of Doppler imaging of HD~11753 obtained from 
observations carried out in January 2010. The elements used are from top to 
bottom Y~II, Cr~II and Ti~II. (from \cite{Korhonen13}) 
{\bf Right)} Measurements of the mean longitudinal magnetic field of HD~11753 
as a function of rotation phase from the January 2010 HARPS data, using Y~II, 
Fe~II, Cr~II and Ti~II line samples separately. Note the correspondence between
the magnetic field polarities and the elemental distributions presented on the 
left. (from \cite{Hubrig12}).}
\label{HD11753}
\end{center}
\end{figure}

A recent study suggests possible intriguing correlations between the magnetic field, abundance anomalies and binary properties in HgMn stars \cite{Hubrig12}. For example in two SB2 systems with synchronously rotating components, the stellar surfaces facing the companion star usually display underabundant element spots and the magnetic field measurements imply negative field polarity. The surface of the opposite hemisphere as a rule is covered by overabundant element spots, and the magnetic field measurements imply positive polarity around the rotation phase of best spot visibility. Similarly, as shown in  Figure\,\ref{HD11753}, in HD~11753 the element underabundance (resp. overabundance) is observed where the magnetic field measurements tend to yield negative (resp. positive) polarity.

On the whole, to answer the question on the existence of the magnetic fields on HgMn stars, more high resolution and high signal-to-noise ratio spectropolarimetric observations are needed.

\section{Final remarks}

HgMn stars are a peculiar group of stars, even among the chemically peculiar stars. For a long time it was thought that inhomogeneous surface abundances would only occur in presence of strong organised magnetic field. The discovery of chemical spots on HgMn stars, which lack this kind of magnetic fields, was a surprise. 

The exact reasons for peculiar chemical abundances and their inhomogeneous surface distributions on HgMn stars are not known. But clues to the answer can be sought from their general properties:  

\begin{itemize}
\item HgMn stars are usually part of a binary or multiple systems
\item They show slower than normal rotation for their spectral type
\item Many of them show line-profile variations caused by chemical spots
\item Spot evolution on timescales of years to months has been reported
\item Even though HgMn stars have spots, they do not have strong large scale magnetic fields. Weak magnetic fields have been reported in some HgMn stars, but more detailed studies of magnetic fields using high resolution, high signal-to-noise spectropolarimetric observations are needed to resolve the issue concerning magnetic field detections.
\end{itemize}

\bigskip
{\it Acknowledgements.} The author acknowledges support by the European Commission under the Marie Curie Intra-European Fellowship Programme in FP7.

\end{document}